\documentclass[dvips]{acta}
\usepackage{supertabular,lscape,epsfig}
\usepackage{amssymb}
\usepackage{amsmath}
\usepackage{wasysym}
\DeclareSymbolFont{ppa}{OT1}{ppl}{m}{it}
\DeclareMathSymbol{\vv}{\mathalpha}{ppa}{'166}

\newfont{\hb}{rphvb at 10pt}%bezszeryfowe pó³grube
\newfont{\hbo}{rphvbo at 10pt}%bezszeryfowe pó³grube kursywa
\newfont{\bitt}{rptmbi at 12pt}%pó³gruba kursywa (tytu³ artyku³u)
\newfont{\bits}{rptmbi at 11pt}%pó³gruba kursywa (tytu³y rozdzia³ów)

\SetPages{285}{301}

\SetVol{61}{2011}

\begin{document}

%Zwarte naglowki, jeden wiersz, appendix
\newcommand{\TabApp}[2]{\begin{center}\parbox[t]{#1}{\centerline{
  {\bf Appendix}}
  \vskip2mm
  \centerline{\small {\spaceskip 2pt plus 1pt minus 1pt T a b l e}
  \refstepcounter{table}\thetable}
  \vskip2mm
  \centerline{\footnotesize #2}}
  \vskip3mm
\end{center}}

%Zwarte naglowki, jeden wiersz
\newcommand{\TabCapp}[2]{\begin{center}\parbox[t]{#1}{\centerline{
  \small {\spaceskip 2pt plus 1pt minus 1pt T a b l e}
  \refstepcounter{table}\thetable}
  \vskip2mm
  \centerline{\footnotesize #2}}
  \vskip3mm
\end{center}}

%Zwarte naglowki, dwa wiersze
\newcommand{\TTabCap}[3]{\begin{center}\parbox[t]{#1}{\centerline{
  \small {\spaceskip 2pt plus 1pt minus 1pt T a b l e}
  \refstepcounter{table}\thetable}
  \vskip2mm
  \centerline{\footnotesize #2}
  \centerline{\footnotesize #3}}
  \vskip1mm
\end{center}}

%Zwarte naglowki, jeden wiersz, appendix
\newcommand{\MakeTableApp}[4]{\begin{table}[p]\TabApp{#2}{#3}
  \begin{center} \TableFont \begin{tabular}{#1} #4 
  \end{tabular}\end{center}\end{table}}

%Zwarte naglowki, jeden wiersz
\newcommand{\MakeTableSepp}[4]{\begin{table}[p]\TabCapp{#2}{#3}
  \begin{center} \TableFont \begin{tabular}{#1} #4 
  \end{tabular}\end{center}\end{table}}

%Zwarte naglowki, jeden wiersz
\newcommand{\MakeTableee}[4]{\begin{table}[htb]\TabCapp{#2}{#3}
  \begin{center} \TableFont \begin{tabular}{#1} #4
  \end{tabular}\end{center}\end{table}}

%Zwarte naglowki, dwa wiersze
\newcommand{\MakeTablee}[5]{\begin{table}[htb]\TTabCap{#2}{#3}{#4}
  \begin{center} \TableFont \begin{tabular}{#1} #5 
  \end{tabular}\end{center}\end{table}}

%FWHM, PSF - proste, MgII, H$\alpha$
%rms, rhs, sd - kursywa
%{\sc DAOPhot}
%{\sf files}
%Galactic wszystko (bulge, center, plane...)
%Cepheids
%type~ Cepheids, Population~II Cepheids
\newfont{\bb}{ptmbi8t at 12pt}
\newfont{\bbb}{cmbxti10}
\newfont{\bbbb}{cmbxti10 at 9pt}
\newcommand{\uprule}{\rule{0pt}{2.5ex}}
\newcommand{\douprule}{\rule[-2ex]{0pt}{4.5ex}}
\newcommand{\dorule}{\rule[-2ex]{0pt}{2ex}}
\def\thefootnote{\fnsymbol{footnote}}

\hyphenation{OGLE}

\begin{Titlepage}
\Title{The Optical Gravitational Lensing Experiment.\\
The OGLE-III Catalog of Variable Stars.\\
XIV.~Classical and Type~II Cepheids in the Galactic Bulge\footnote{Based on
observations obtained with the 1.3-m Warsaw telescope at the Las Campanas
Observatory of the Carnegie Institution for Science.}}
\vspace*{5pt}
\Author{I.~~S~o~s~z~y~ñ~s~k~i$^1$,~~
A.~~U~d~a~l~s~k~i$^1$,~~
P.~~P~i~e~t~r~u~k~o~w~i~c~z$^1$,~~
M.\,K.~~S~z~y~m~a~ñ~s~k~i$^1$,\\
M.~~K~u~b~i~a~k$^1$,~~
G.~~P~i~e~t~r~z~y~ñ~s~k~i$^{1,2}$,~~
£.~~W~y~r~z~y~k~o~w~s~k~i$^{1,3}$,~~
K.~~U~l~a~c~z~y~k$^1$,~~\\
R.~~P~o~l~e~s~k~i$^1$~~
and~~S.~~K~o~z~³~o~w~s~k~i$^1$}
{$^1$Warsaw University Observatory, Al.~Ujazdowskie~4, 00-478~Warszawa, Poland\\
e-mail:
(soszynsk,udalski,pietruk,msz,mk,pietrzyn,kulaczyk,rpoleski,simkoz)@astrouw.edu.pl\\
$^2$Universidad de Concepción, Departamento de Astronomia, Casilla 160--C, Concepción, Chile\\
$^3$Institute of Astronomy, University of Cambridge, Madingley Road, Cambridge CB3~0HA,~UK\\
e-mail: wyrzykow@ast.cam.ac.uk}
\vspace*{-5pt}
\Received{November 21, 2011}
\end{Titlepage}
\vspace*{-5pt}
\Abstract{The fourteenth part of the OGLE-III Catalog of Variable Stars
(OIII-CVS) contains Cepheid variables detected in the OGLE-II and OGLE-III
fields toward the Galactic bulge. The catalog is divided into two main
categories: 32 classical Cepheids (21 single-mode fundamental-mode F, four
first-overtone 1O, two double-mode F/1O, three double-mode 1O/2O and two
triple-mode 1O/2O/3O pulsators) and 335 type~II Cepheids (156 BL~Her, 128
W~Vir and 51 RV~Tau stars). Six of the type~II Cepheids likely belong to
the Sagittarius Dwarf Spheroidal Galaxy. The catalog data include the
time-series photometry collected in the course of the OGLE survey,
observational parameters of the stars, finding charts, and
cross-identifications with the General Catalogue of Variable Stars.

We discuss some statistical properties of the sample and compare it with
the OGLE catalogs of Cepheids in the Large and Small Magellanic Clouds.
Multi-mode classical Cepheids in the Galactic bulge show systematically
smaller period ratios than their counterparts in the Magellanic Clouds.
BL~Her in the Galactic bulge stars seem to be brighter than the linear
extension of the period--luminosity relations defined by the longer-period
type~II Cepheids. We also show individual stars of particular interest,
like two BL~Her stars with period doubling.}{Stars: variables: Cepheids
-- Stars: oscillations (including pulsations) -- Stars: Population II --
Galaxy: center}

\Section{Introduction}
The OGLE-III Catalog of Variable Stars (OIII-CVS) is a continuously growing
collection of variable stars detected in the fields observed in the course
of the third phase of the Optical Gravitational Lensing Experiment. The
parts of the catalog published so far contain in total 193\,000 variable
stars of various types identified in the Large (LMC) and Small Magellanic
Clouds (SMC) and in the Galactic bulge. Among others, we published the
catalogs of Cepheids in the LMC (Soszyñski \etal 2008ab) and SMC (Soszyñski
\etal 2010ab), and recently we released the catalog of 16\,836 RR~Lyr stars
in the Galactic center (Soszyñski \etal 2011). This paper presents the
sample of classical and type~II Cepheids detected in the 68.7~square
degrees toward the Galactic bulge.

The first Cepheids in the direction of the Galactic center (but located
much closer, in the foreground of the bulge) were discovered in the
nineteenth century (Schmidt 1866, Sawyer 1886, Roberts 1894). The first
Cepheids that belong to the bulge were discovered by A.~J.~Cannon during
the survey for variable stars carried out by the Harvard College
Observatory (Pickering 1908, 1911). In subsequent decades, the effort of
many observers (\eg Swope 1940, Hertzsprung 1941, Plaut 1948, 1958, 1971,
Oosterhoff and Horikx 1952, Oosterhoff \etal 1954, Ponsen 1955, Hoffleit
1964, Kooreman 1968) increased to about 150 the number of known classical
and type~II Cepheids in the region of the Galactic bulge.

Since the late twentieth century the number of known variable stars has
grown very fast thanks to the huge amount of photometric data collected by
the microlensing surveys. The second phase of the OGLE project (OGLE-II)
have yielded the sample of 54 type~II Cepheids in the Galactic bulge
presented by Kubiak and Udalski (2003). This sample, extended by several
Cepheids from the OGLE-III project, was analyzed by Groenewegen \etal
(2008). From the latest studies, it is worth noting the discovery of three
long-period classical Cepheids in the nuclear bulge done in the
near-infrared domain by Matsunaga \etal (2011b). This discovery affects our
understanding of the star formation history in the center of the Milky Way.

The catalog presented here contains both, classical and type~II, Cepheids
observed toward the Galactic bulge. Classical Cepheids in our sample are
generally short-period variables and some of them seem to be located behind
the Galactic bulge. Variables of this type in front of the bulge are too
bright and saturate in the OGLE images. Most of our type~II Cepheids belong
to the Galactic bulge, which results in a distinct period--luminosity (PL)
relation followed by these stars. Several Cepheids in our catalog likely
belong to the Sagittarius Dwarf Spheroidal Galaxy (Sgr dSph).

The paper is structured as follows. In Section~2 we present the photometric
data and the reduction methods. Section~3 describes the selection and
classification of the Cepheids. In Section~4 we present the catalog itself
and compare our sample with the set of variables included in the General
Catalogue of Variable Stars (GCVS) and other sources. In Section~5 we
discuss our results.
\vspace*{-5pt}
\Section{Observational Data}
\vspace*{-3pt}
The catalog was compiled from the observations taken between 1997 and 2009
by the second and third stages of the OGLE survey. Both stages were carried
out with the 1.3-m Warsaw telescope at the Las Campanas Observatory,
Chile. The observatory is operated by the Carnegie Institution for
Science. During the OGLE-II project (1997--2000) the Warsaw telescope was
equipped with the ``first generation'' camera with a SITe $2048\times2048$
CCD detector working in drift-scan mode. In 2001 this camera was replaced
by the eight chip mosaic CCD camera (Udalski 2003a) covering approximately
$35\times35$~arcmin$^2$ in the sky with the scale of 0.26 arcsec/pixel.

Time-series photometry was obtained with the {\it I} and {\it V} filters,
closely resembling the standard system. The number of observations
significantly varies from field to field: from several dozen to about 3000
measurements in the {\it I}-band, and up to 50 points in the {\it
V}-band. Similarly to the catalog of RR~Lyr stars (Soszyñski \etal 2011) we
restricted our search only to the fields with at least 30 {\it I}-band
observations, which gave the sky coverage of 68.7~square degrees.

The OGLE data reduction pipeline is based on the Difference Image Analysis
(DIA) method developed by Alard and Lupton (1998) and Wo¼niak (2000). For
18 bright Cepheids, which saturate in the DIA reference images, we provide
the {\sc DoPhot} {\it I}-band photometry (Schechter \etal 1993). These
stars are flagged in the remarks of the catalog. Instrumental magnitudes
were transformed to the standard system using the procedure introduced by
Udalski \etal (2008). Following the recipe of Szymañski \etal (2011) we
also applied a small correction to the {\it I}-band measurements for stars
redder than $(V-I)=1.5$~mag.

Since many Cepheids in our catalog have a limited number of observations in
the {\it V}-band, we used the following method to derive the $(V-I)$ color
indices for these objects. For a given star we used its well sampled {\it
I}-band light curve as a template light curve which was fitted to the {\it
V}-band observations. For each such a template light curve we adjusted the
mean magnitude, amplitude and phase shift between {\it I}- and {\it
V}-bands. Then using our transformed template light curve we derived
intensity mean magnitude in the {\it V}-band.

The magnitudes of Cepheids were dereddened using the extinction maps
obtained by Pietrukowicz \etal (2011) on the basis of the OGLE-III catalog
of RR~Lyr stars (Soszyñski \etal 2011). Following Pietrukowicz \etal (2011)
we adopted an anomalous extinction law, $R_I=A_I/E(V-I)=1.08$, in agreement
with the earlier investigation of Udalski (2003b).

\vspace*{-5pt}
\Section{Selection and Classification of Cepheids}
\vspace*{-3pt}
The selection of Cepheids toward the Galactic bulge was performed using
generally the same methods as for RR~Lyr stars (Soszyñski \etal 2011). We
searched the database of over $3\times10^8$ stars observed by the OGLE
project around the Milky Way center. For each star we found the most
significant periods using the {\sc Fnpeaks} code kindly provided by
Z.\,Ko³aczkowski. The {\it I}-band light curves with the primary periods
between 1 and 50~days and amplitudes above 0.1~mag were inspected by eye
and initially divided into pulsating, eclipsing and other variable stars.

Most of the ``other'' variables turned out to be spotted stars with
characteristic changes of amplitude, mean luminosity and light curve
shape. The light curves of some of these objects are very similar to the
pulsating stars (especially when the OGLE observations covered only two or
three seasons) and were initially categorized as Cepheids. However, upon
further analysis we investigated the OGLE-IV light curves of our candidates
for pulsating stars and we filtered out the spotted variables from our
catalog. The final pulsation periods given in this catalog were derived
with the {\sc Tatry} code (Schwarzenberg-Czerny 1996). Each Cepheid was
also searched for secondary periods by prewhitening the light curves with
the primary periods and deriving periodicities for residual data.

Our final classification of variable stars was based primarily on their
light curve morphology, but we also took into account their positions in
the PL and color--magnitude diagrams, ratio of amplitudes in the {\it I}-
and {\it V}-bands, period ratios (for multiperiodic stars), etc. Our final
list of Cepheids toward the Galactic bulge contains 367 objects.

\Subsection{Classical Cepheids}
When Baade (1952) announced his revision of the extragalactic distance
scale, it became clear that classical (or type~I) and type~II Cepheids
constitute separate classes of variable stars. Our catalog contains 32
classical Cepheids, of which 11 stars have the dominant pulsation periods
below 1~day and were identified by us earlier, during the search for RR~Lyr
stars in the Galactic bulge (Soszyñski \etal 2011). Actually, several of
these short-period Cepheids were included in the catalog of RR~Lyr stars,
usually with a remark ``possible Cepheid''. We provide their
cross-identifications in the remarks of the catalog.

\begin{figure}[htb]
\centerline{\includegraphics[width=12.2cm, bb=65 450 555 755]{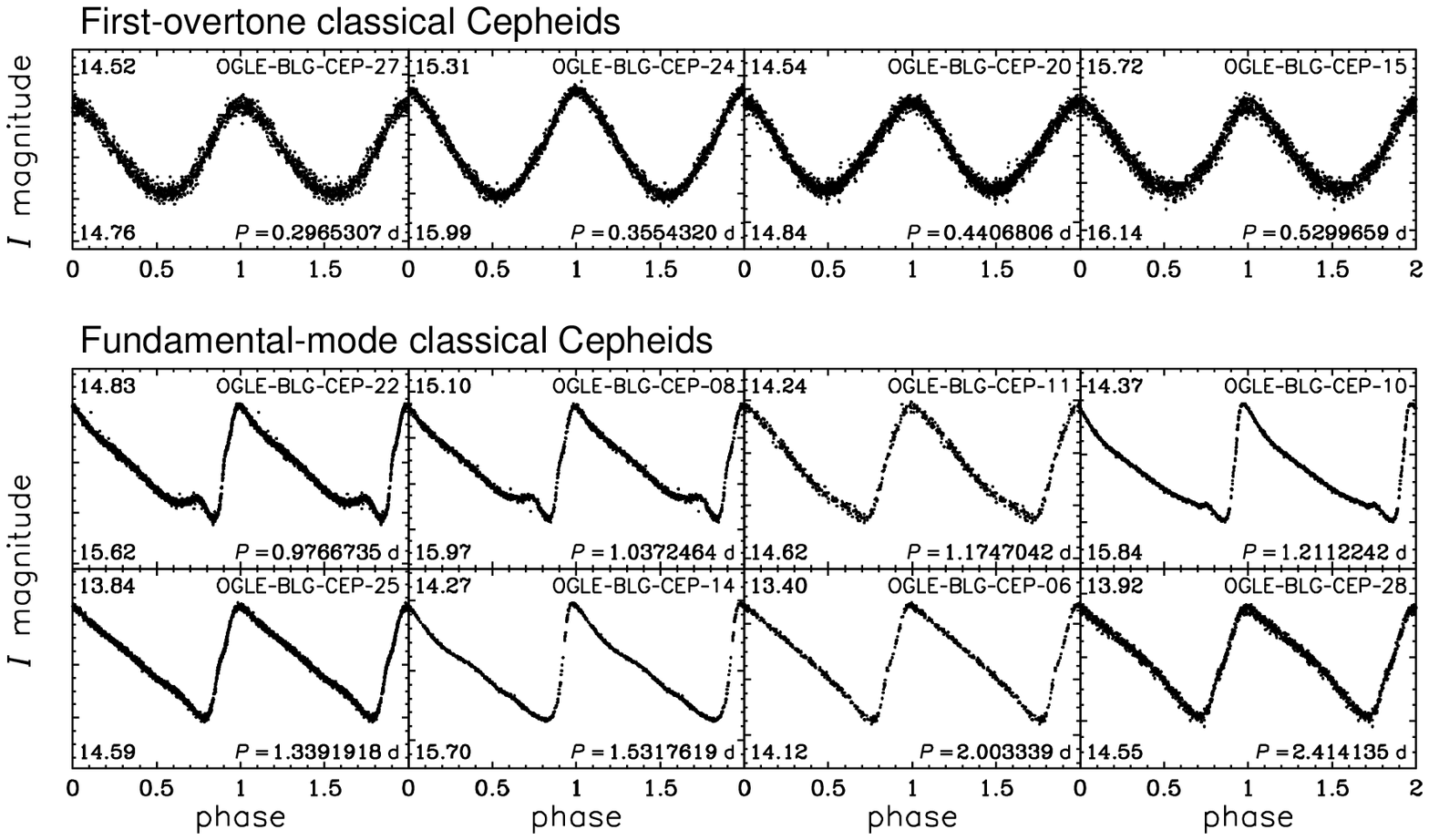}}
\FigCap{Sample light curves of single-mode classical Cepheids in the
Galactic bulge. {\it Upper panels} show first-overtone stars, while {\it
lower panels} present fundamental-mode Cepheids brighter than the PL
relation for type~II Cepheids (see discussion in Section~5). Note that the
range of magnitudes varies from panel to panel. Numbers in the left corners
show the minimum and maximum magnitudes of the range.}
\end{figure}

The OGLE-III catalogs of pulsating stars in the LMC (Soszyñski \etal 2008a,
Poleski \etal 2010) revealed that the first-overtone classical Cepheids and
$\delta$~Sct stars follow the same, continuous PL relation, while the
fundamental-mode pulsators are naturally divided by the lack of stars with
periods between approximately 0.4 and 1~day. Thus, the boundary period
between the overtone Cepheids and $\delta$~Sct stars may be freely adopted,
and we used a similar borderline like in the Magellanic Clouds. The
shortest first-overtone period in our catalog is 0.23~days. The
fundamental-mode Cepheids toward the bulge have periods longer than
0.97~days, with the exception of a double-mode object OGLE-BLG-CEP-21 with
the funda\-mental-mode period of about 0.78~days. In the OGLE database we
also detected double- and triple-mode pulsators with the fundamental-mode
periods of about 0.4~days. We classified these objects as $\delta$~Sct
stars and we will include them in the next parts of our Catalog.

Four short-period classical Cepheids in our sample are single-mode overtone
pulsators. The light curves of these stars are shown in Fig.~1. We also
found two double-mode Cepheids pulsating simultaneously in the fundamental
mode and the first overtone, three pulsators with the first and second
overtones excited and two triple-mode Cepheids with the first three
overtones excited. The remaining 21 classical Cepheids are fundamental-mode
pulsators, which were distinguished from the type~II Cepheids on the basis
of their light curve shapes.

\begin{figure}[htb]
\centerline{\includegraphics[width=12.9cm]{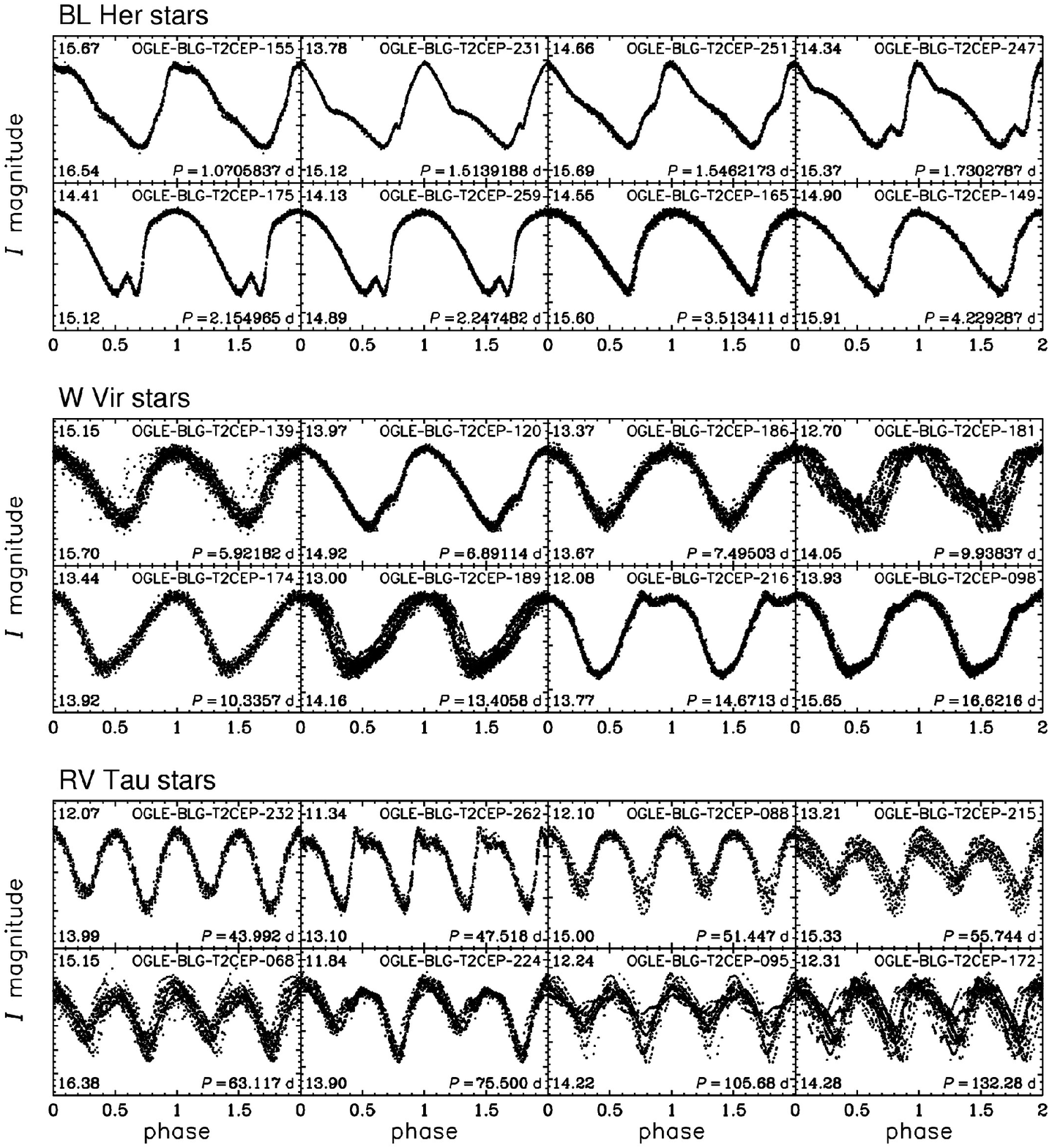}}
\FigCap{Sample {\it I}-band light curves of type~II Cepheids in the
Galactic bulge. {\it Upper panels} show BL~Her stars, {\it middle panels}
-- W~Vir stars, and {\it lower panels} -- RV~Tau stars (folded with the
``double'' periods). Note that the range of magnitudes varies from panel to
panel. Numbers in the left corners show the minimum and maximum magnitudes
of the range.}
\end{figure}

\Subsection{Type~II Cepheids}
Type~II Cepheids (sometimes called Population~II Cepheids) are low-mass
pulsating stars of the thick disk and halo populations. They are divided
into three subtypes -- BL~Her, W~Vir, and RV~Tau stars -- which are
associated with consecutive phases of stellar evolution after exhaustion of
helium in the core (Schwarzschild and H{\"a}rm 1970, Strom \etal 1970,
Gingold 1974, 1976). BL~Her stars are post-horizontal branch stars evolving
towards the asymptotic giant branch. W~Vir stars are thought to undergo
blueward loops from the asymptotic giant branch due to helium shell
flashes. RV~Tau stars cross the instability strip evolving away from the
asymptotic giant branch toward the white dwarf domain.

\begin{figure}[tb]
\centerline{\includegraphics[width=11.6cm, bb=65 125 555 745]{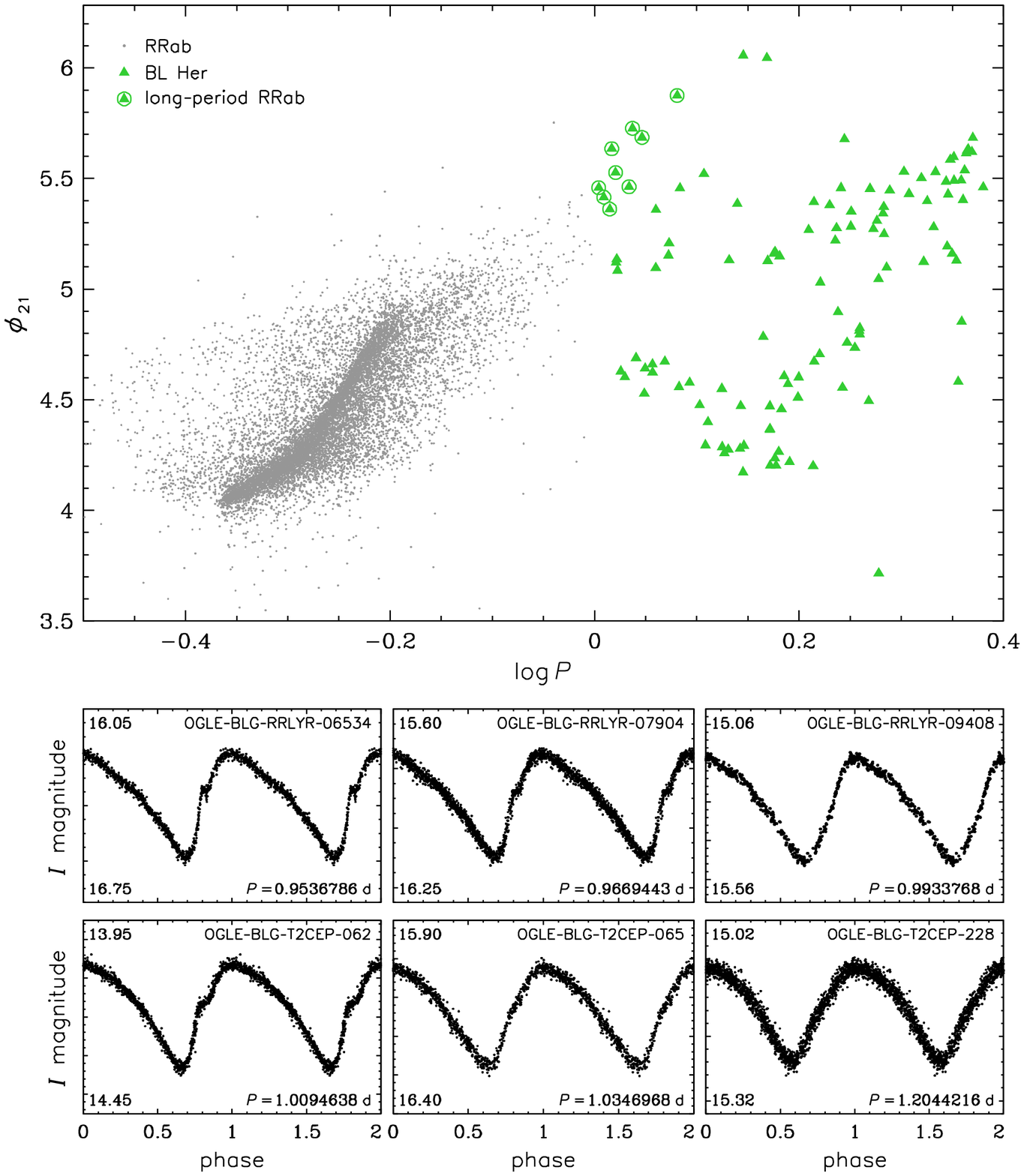}}
\FigCap{{\it Upper panel}: Fourier coefficients $\phi_{21}$ \vs the
logarithm of the periods for RRab stars (grey points) and short-period
type~II Cepheids (green triangles). Candidates for RR~Lyr stars with
periods above 1~day are surrounded by circles. {\it Lower panels}: Light
curves of three RR~Lyr stars with periods just below 1~day, and three
candidates for long-period RR~Lyr stars.}
\end{figure}

Our catalog contains 335 type~II Cepheids. Fig.~2 presents an example set
of light curves from the sample. As in the catalogs in the LMC and SMC
(Soszyñski \etal 2008b, 2010b), the transition between BL~Her and RR~Lyr
stars was defined at $P=1$~day. However, the parameters of the Fourier
decomposition of light curves suggest that the two classes overlap in
period space and the distinction between them is not a clear-cut. Fig.~3
shows the Fourier coefficient $\phi_{21}$ (Simon and Lee 1981) plotted
against the logarithm of the period for RRab stars (from the catalog of
Soszyñski \etal 2011) and short-period BL~Her stars. The symbols surrounded
by circles indicate nine BL~Her stars which seem to be an extension of the
RR~Lyr stars to periods longer than 1~day. Lower panels of Fig.~3 present
light curves of three typical RR~Lyr stars with periods just below 1~day,
and three stars with periods above 1~day, but likely belonging to the same
class. These objects may be called long-period RR~Lyr stars, which moves
the upper limit of the RR~Lyr periods to over 1.2~days. However, as we
mentioned, in this catalog all Population~II pulsators with periods longer
than 1~day are called BL~Her stars.

On the other hand, one can expect that period of 1~day is not the shortest
period of the genuine BL~Her stars and some type~II Cepheids may have been
included in our catalog of RR~Lyr stars in the bulge (Soszyñski \etal
2011). The light curves of the genuine BL~Her stars are shown in
Fig.~2. Their Fourier coefficient $\phi_{21}$ is below 4.7 for periods of
about 1~day. The stars with similar periods and with $4.7<\phi_{21}<5.3$,
\ie located between genuine BL~Her and long-period RR~Lyr stars, may be
included to both groups. It is noteworthy that in the Magellanic Clouds we
did not notice such a distinct group of long-period RR~Lyr stars. An
attempt to separate long-period RR~Lyr stars from the short-period BL~Her
stars based on their light curves was undertaken by Diethelm (1983).

\begin{figure}[tb]
\centerline{\includegraphics[width=11.6cm, bb=65 415 555 745]{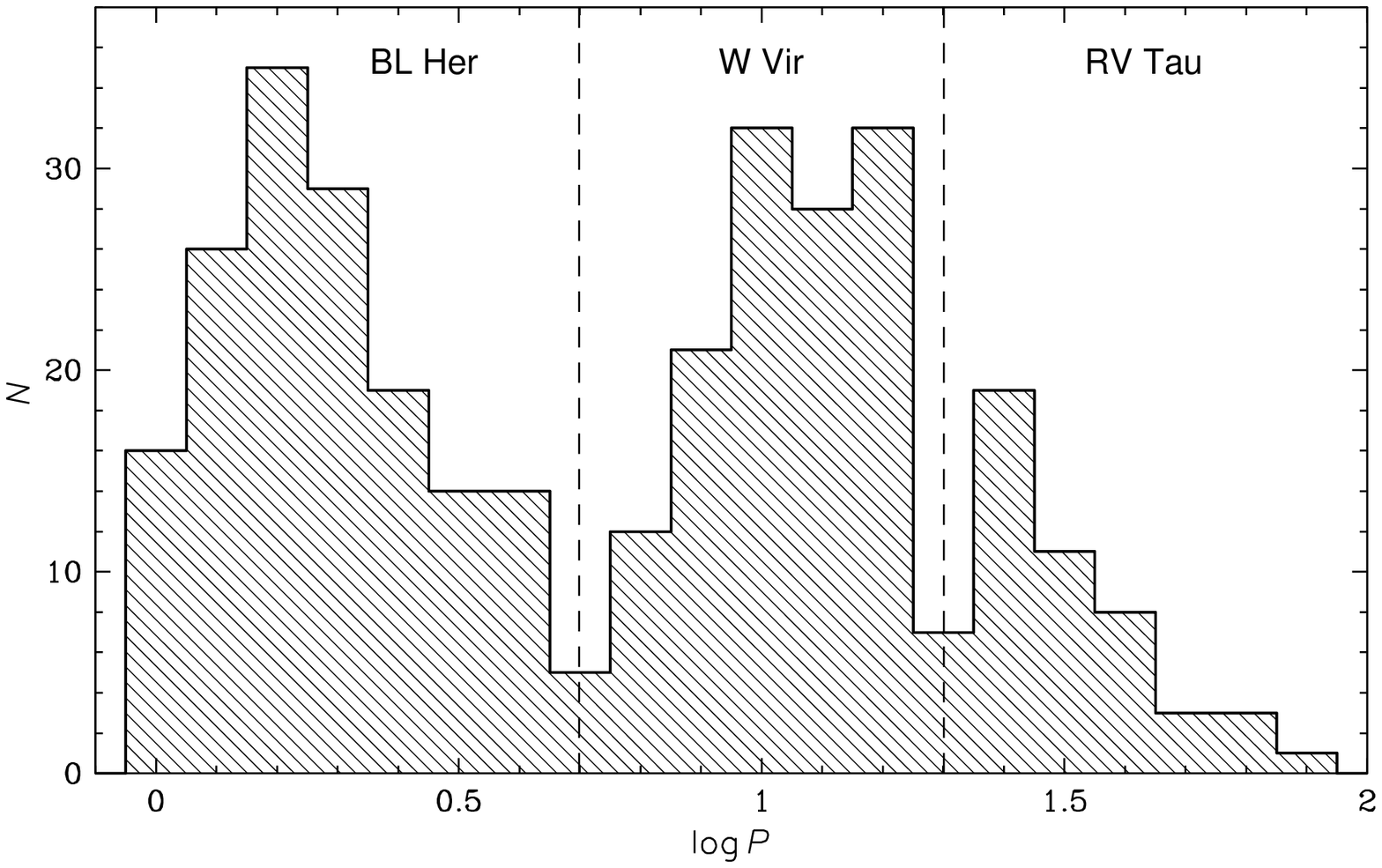}}
\FigCap{Period distribution of type~II Cepheids in the Galactic
bulge. Vertical dashed lines show the thresholds, 5~d and 20~d, used to
divide BL~Her, W~Vir and RV~Tau stars.}
\end{figure}

As a borderline between BL~Her and W~Vir stars we adopted a period of
5~days, \ie longer than in the Magellanic Clouds (4~days). In the bulge
this value seems to best separate both classes of type~II Cepheids. First,
the period distribution (Fig.~4) shows a local minimum for
$P\approx5$~days. Second, type~II Cepheids with periods below 5~days are on
average bluer than Cepheids with longer periods, which is visible in the
color--magnitude diagram (Fig.~5). Such a discontinuity in $(V-I)$ colors
between BL~Her and V Wir stars was also found in the Magellanic
Clouds. Third, type~II Cepheids with periods above and below 5~days seem to
obey different PL relations. We will return to this point in the
discussion.

\begin{figure}[tb]
\centerline{\includegraphics[width=11.6cm, bb=65 375 555 745]{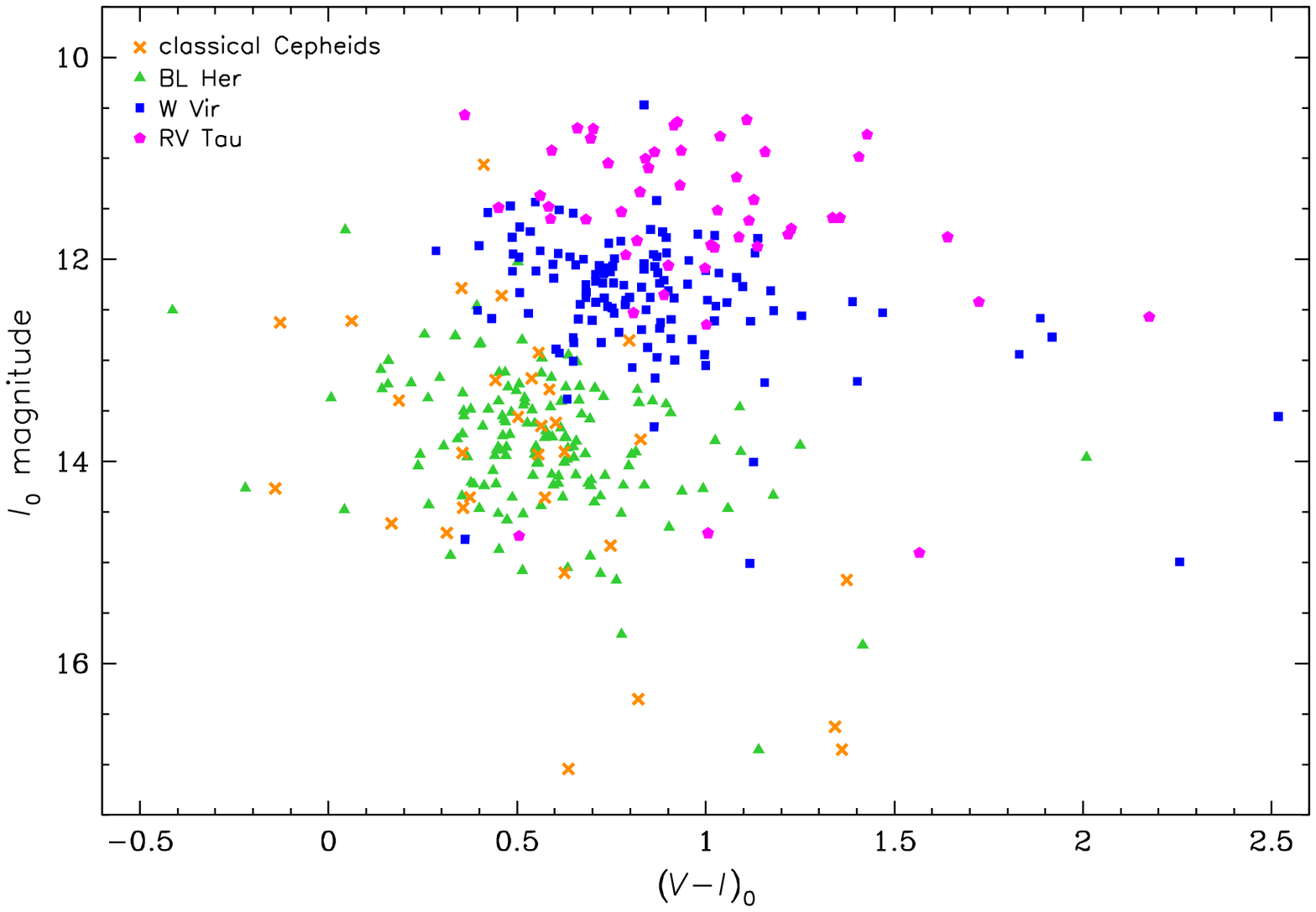}}
\FigCap{Color--magnitude diagram for classical (yellow crosses) and type~II
(green, blue and magenta symbols) Cepheids toward the Galactic bulge. The
color indices and luminosities were dereddened using the extinction maps by
Pietrukowicz \etal (2011).}
\end{figure}

The boundary between W~Vir and RV~Tau stars was traditionally adopted at
$P=20$~days (``single'' period, \ie the interval between successive
minima). The only exception is OGLE-BLG-T2CEP-045 with a single period of
19.36~days, which displays clear alternations of deep and shallow minima,
\ie the typical behavior of RV~Tau stars. Not all long-period type~II
Cepheids ($P>20$~days) show clear alternations of cycles, however all these
stars are called RV~Tau stars in our catalog.

In the Magellanic Clouds (Soszyñski \etal 2008b, 2010b) we distinguished an
additional subclass of type~II Cepheids, which we called peculiar W~Vir
stars. These Cepheids are characterized by higher luminosities, on average
bluer colors and a different light curve morphology than regular W~Vir
stars. Large fraction of peculiar W~Vir stars show eclipsing or ellipsoidal
modulation superimposed on the pulsation light curves, which suggests that
all these stars are members of binary systems. Candidates for peculiar
W~Vir stars were also found in the Milky Way (Matsunaga \etal 2009, Welch
and Foster 2011).

In the Galactic bulge we identified no evident peculiar W~Vir
star. Although a lot of type~II Cepheids have Fourier coefficients similar
to the peculiar W~Vir stars, these objects do not satisfy other conditions:
they are not brighter and bluer than regular W~Vir stars. We also did not
detect any Cepheid with signs of binarity. Thus, in this catalog we did not
divided W~Vir stars into ``regular'' and ``peculiar'' variables.

\Section{Catalog of Cepheids Toward the Galactic Bulge}
The OGLE-III catalog of classical Cepheids toward the Galactic bulge
comprises 32 objects, of which 21 pulsate solely in the fundamental mode
(F), four are single-mode first-overtone pulsators (1O), two stars are F/1O
double-mode Cepheids, three -- 1O/2O double-mode Cepheids, and two --
1O/2O/3O triple-mode Cepheids. The sample of 335 type~II Cepheids has been
divided into 156 BL~Her, 128 W~Vir, and 51 RV~Tau stars. Six type~II
Cepheids likely belong to Sgr dSph.

The catalogs are accessible through the anonymous FTP site or {\it via} the
web interface:
\vspace*{-10pt}
\begin{center}
{\it ftp://ftp.astrouw.edu.pl/ogle/ogle3/OIII-CVS/blg/cep/}\\
{\it ftp://ftp.astrouw.edu.pl/ogle/ogle3/OIII-CVS/blg/t2cep/}\\
{\it http://ogle.astrouw.edu.pl}\\
\end{center}
\vspace*{-4pt}
The FTP sites are organized as follows. The lists of Cepheids with their
J2000 equatorial coordinates, classification, identifications in the
OGLE-II and OGLE-III databases and in the GCVS are given in the {\sf
ident.dat} files. The stars are arranged in order of increasing right
ascension and designated OGLE-BLG-CEP-NN or OGLE-BLG-T2CEP-NNN for
classical and type~II Cepheids, respectively. The observational parameters
of the Cepheids -- intensity-averaged $\langle I\rangle$ and $\langle
V\rangle$ magnitudes, periods with uncertainties (derived with the {\sc
Tatry} code by Schwarzenberg-Czerny 1996), peak-to-peak {\it I}-band
amplitudes and parameters of the Fourier light curve decomposition
$R_{21}$, $\phi_{21}$, $R_{31}$, $\phi_{31}$ (Simon and Lee 1981) -- are
given in separate files. Additional information on some objects can be
found in the files {\sf remarks.txt}. The OGLE-II and OGLE-III multi-epoch
{\it VI} photometry can be downloaded from the directory {\sf
phot/}. Finding charts for each star are stored in the directory {\sf
fcharts/}. These are $60\arcs\times60\arcs$ subframes of the {\it I}-band
DIA reference images.

The completeness of the catalog was judged by comparing our sample with the
list of type~II Cepheids found by Kubiak and Udalski (2003) and with the
GCVS (Samus \etal 2011). The cross-identification with the OGLE-II catalog
of Kubiak and Udalski (2003) revealed 12 missing objects in our sample, of
which eight stars have periods below 1~day and were included in the
OGLE-III catalog of RR~Lyr stars (Soszyñski \etal 2011). The remaining four
missing stars turned out to be reclassified in the OGLE-III database as
eclipsing or spotted variables.

We cross-matched our catalog with the stars classified as CEP, DCEP, CW,
CWA, CWB, RV, RVA or RVB in the GCVS. From 49 such objects, that can be
potentially found in the OGLE-II and OGLE-III fields in the Galactic bulge,
we unambiguously identified 24 Cepheids. Most of the remaining objects
turned out to be much too bright to be present in the OGLE database. Some
variables classified in the GCVS as RV~Tau stars have light curves
indistinguishable from the long-period variables (SRV). These objects also
have not been included in our catalog. From the six type~II Cepheids
discovered with the Hubble Space Telescope by Pritzl \etal (2003) in the
core of the globular cluster NGC~6441 we detected only one object (V6 =
OGLE-BLG-T2CEP-083) -- the most distant one from the center of the cluster.

It seems that the OGLE-III catalog of Cepheids in the Galactic bulge is
nearly complete for stars with the {\it I}-band magnitudes in the range
12--20~mag, but there are brighter and probably fainter Cepheids which are
not included in our sample. Also, our catalog is incomplete in the cores of
dense globular clusters.

\begin{figure}[tb]
\centerline{\includegraphics[width=12cm, bb=55 50 555 745]{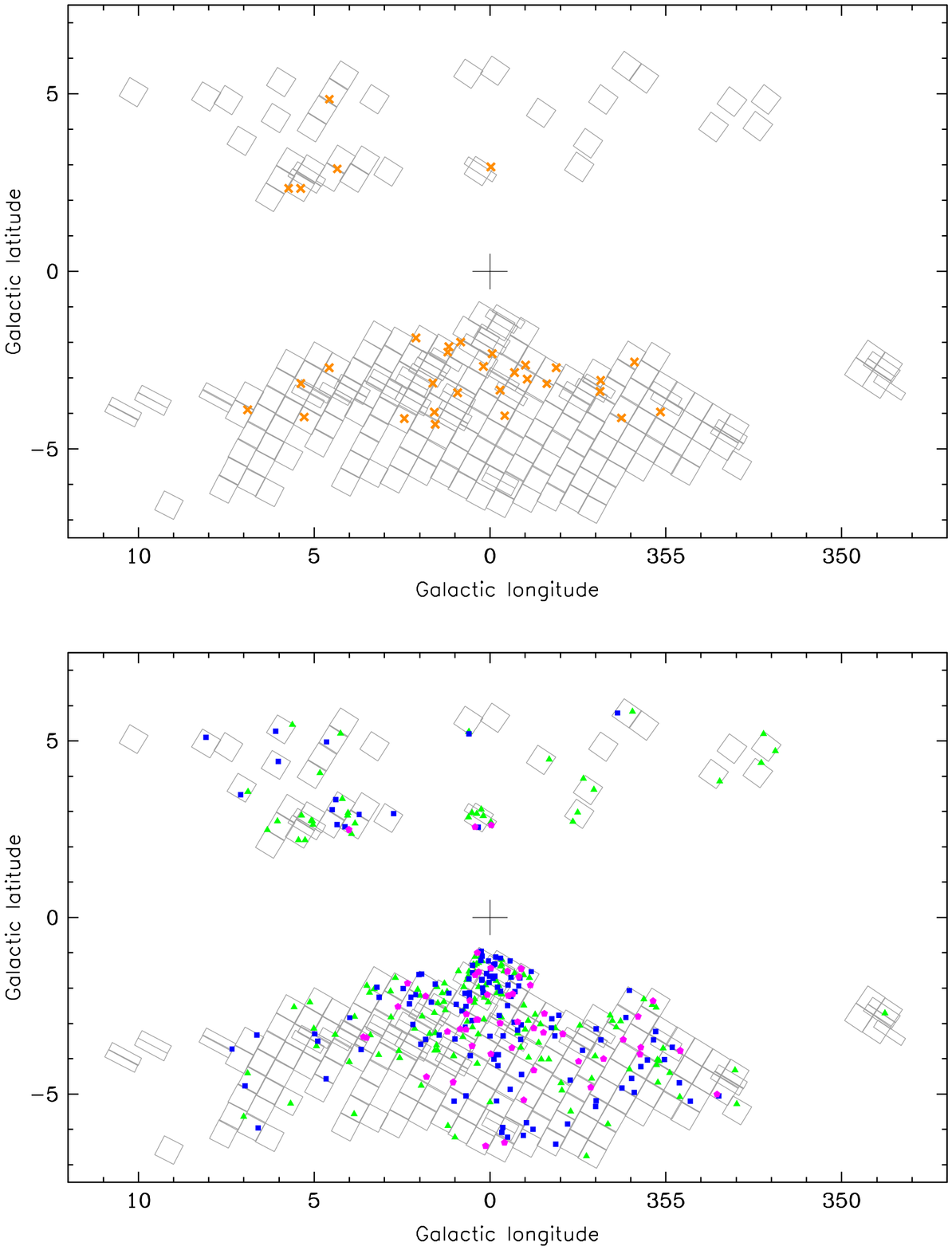}}
\FigCap{Spatial distribution of classical ({\it upper panel}) and type~II
({\it lower panel}) Cepheids toward the Galactic bulge. Different symbols
correspond to types of stars as shown in Fig.~5. Grey contours show the
OGLE-II and OGLE-III fields with the number of observations exceeding 30.}
\end{figure}

Fig.~6 shows the position of classical and type~II Cepheids in the sky. One
can notice different spatial distributions of both types of Cepheids. Old
stars are strongly concentrated toward the Galactic center, like RR~Lyr
stars (Soszyñski \etal 2011). Classical Cepheids are distributed almost
parallel to the Galactic plane, between Galactic latitudes
$-5\arcd<b<5\arcd$. The OGLE-III fields closest to the Galactic center
(with the largest interstellar extinction) are completely omitted by the
classical Cepheids.

\vspace*{-9pt}
\Section{Discussion}
\vspace*{-5pt}
\Subsection{Classical Cepheids}
\vspace*{-4pt}
It has been known for a long time that in the Galactic bulge type~II
Cepheids dominate over classical Cepheids (Oosterhoff 1956). Indeed, our
catalog contains an order of magnitude larger sample of type~II Cepheids
than classical Cepheids. Moreover, the reddening-free PL diagram shown in
Fig.~7 raises doubts whether the observed classical Cepheids belong to the
bulge.
\begin{figure}[htb]
\centerline{\includegraphics[width=11.7cm, bb=55 375 555 755]{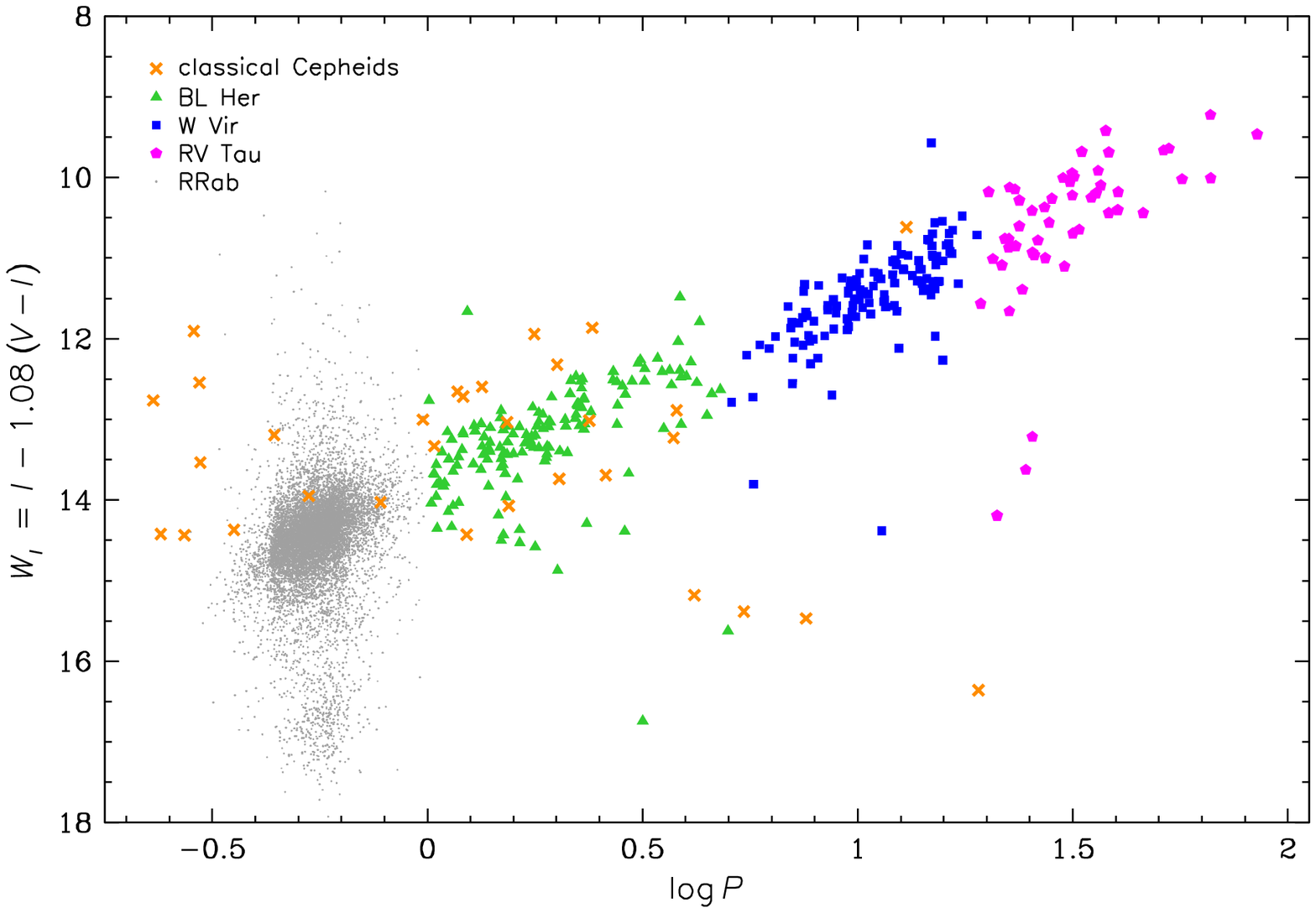}}
\FigCap{Period--$W_I$ diagram for classical Cepheids (yellow crosses),
type~II Cepheids (green, blue and magenta symbols) and RR~Lyr stars (grey
points) toward the Galactic bulge. $W_I$ is the reddening-free Wesenheit
index, defined as $W_I=I-1.08(V-I)$.}
\end{figure}

While type~II Cepheids concentrate along a well-defined PL relation,
classical Cepheids do not follow a distinct PL sequence. In the LMC
(Soszyñski \etal 2008b) fundamental-mode type I Cepheids are by about
1.3~mag brighter than type~II Cepheids for periods around 1~day, and
2.5~mag brighter for periods above 20~days. In the bulge we have a group of
short-period (0.98--2.41~days) fundamental-mode classical Cepheids by only
0.7~mag brighter than BL~Her stars. The light curves of these variables are
shown in Fig.~1.

There are three possible explanations of this inconsistency between the
Galactic bulge and LMC.
\begin{enumerate}
\item All fundamental-mode classical Cepheids in our sample are located in
the Galactic disk behind the bulge, and therefore they have lower apparent
luminosities compared to type~II Cepheids. This hypothesis is supported by
the spatial distribution of classical Cepheids (Fig.~6), which is
completely different than for type~II Cepheids and other members of the
bulge (RR~Lyr stars, red clump stars). On the other hand, most of the
classical Cepheids in our catalog have colors typical for Cepheids in the
bulge (Fig.~5) and are not additionally reddened by the interstellar matter
behind the bulge. This argument supports the following two hypotheses.
\item The Cepheids above the type~II Cepheids PL relation belong to the
bulge, but have lower absolute luminosities than classical Cepheids. These
objects may be so called anomalous Cepheids -- brighter than type~II and
fainter than classical Cepheids. We found 62 fundamental-mode anomalous
Cepheids in the LMC (Soszyñski \etal 2008b) and only three candidates for
such stars in the SMC (Soszyñski \etal 2010a).
\item The Cepheids from the mentioned group are normal classical Cepheids
belonging to the bulge, but the BL~Her stars in the bulge are
systematically brighter than the same stars in the LMC. Indeed, in Fig.~7
BL~Her stars seem to lie above the linear extension of the PL relation
defined by the longer-period type~II Cepheids. In the SMC BL~Her stars
are also placed above the PL relation fitted to the W~Vir stars
(Matsunaga 2011a).
\end{enumerate}

If the latter possibility occurs, the classical Cepheids in the Galactic
bulge have short periods, below 2.5~days. It is in agreement with Matsunaga
\etal (2011b), who discovered three Cepheids with periods of approximately
20~days, and no Cepheids with shorter periods. However, due to the
detection limits in the near-infrared bands, their survey was sensitive
only to the Cepheids with periods longer than 5~days. Regardless of which
hypothesis is true, there is no doubt that our catalog contains also
classical Cepheids located behind the bulge. They are much fainter and
redder (due to the additional extinction) than typical Cepheids in the
bulge.

The OGLE-III catalog of classical Cepheids toward the Galactic bulge
comprises an exceptionally large fraction (22\% of the total sample) of
multi-mode pulsators. In the LMC and SMC we observed 8\% and 6\% of
multi-mode Cepheids, respectively. The observational parameters of the two
triple-mode Cepheids in the bulge are given in Table~1. Fig.~8 shows their
light curves folded with the three pulsation periods, each one after
prewhitening with the other two modes.

\tabcolsep=2.5pt
\MakeTableee{lcccccccc}{12.5cm}{Triple-mode Cepheids in the Galactic bulge}
{\hline
\noalign{\vskip3pt}
\multicolumn{1}{c}{Star name} & $\langle{I}\rangle$ & $\langle{V}\rangle$ & $P_{\rm 1O}$ & $A_I^{\rm 1O}$ & $P_{\rm 2O}$ & $A_I^{\rm 2O}$ & $P_{\rm 3O}$ & $A_I^{\rm 3O}$ \\
& [mag] & [mag] & [days] & [mag] & [days] & [mag] & [days] & [mag] \\
\noalign{\vskip3pt}
\hline
\noalign{\vskip3pt}
OGLE-BLG-CEP-16 & 13.994 & 15.338 & 0.2954573 & 0.147 & 0.2339744 & 0.014 & 0.1950696 & 0.015 \\
OGLE-BLG-CEP-30 & 14.184 & 15.499 & 0.2303703 & 0.111 & 0.1829840 & 0.027 & 0.1521731 & 0.034 \\
\noalign{\vskip2pt}
\hline}

\begin{figure}[htb]
\centerline{\includegraphics[width=12.2cm, bb=55 465 555 755]{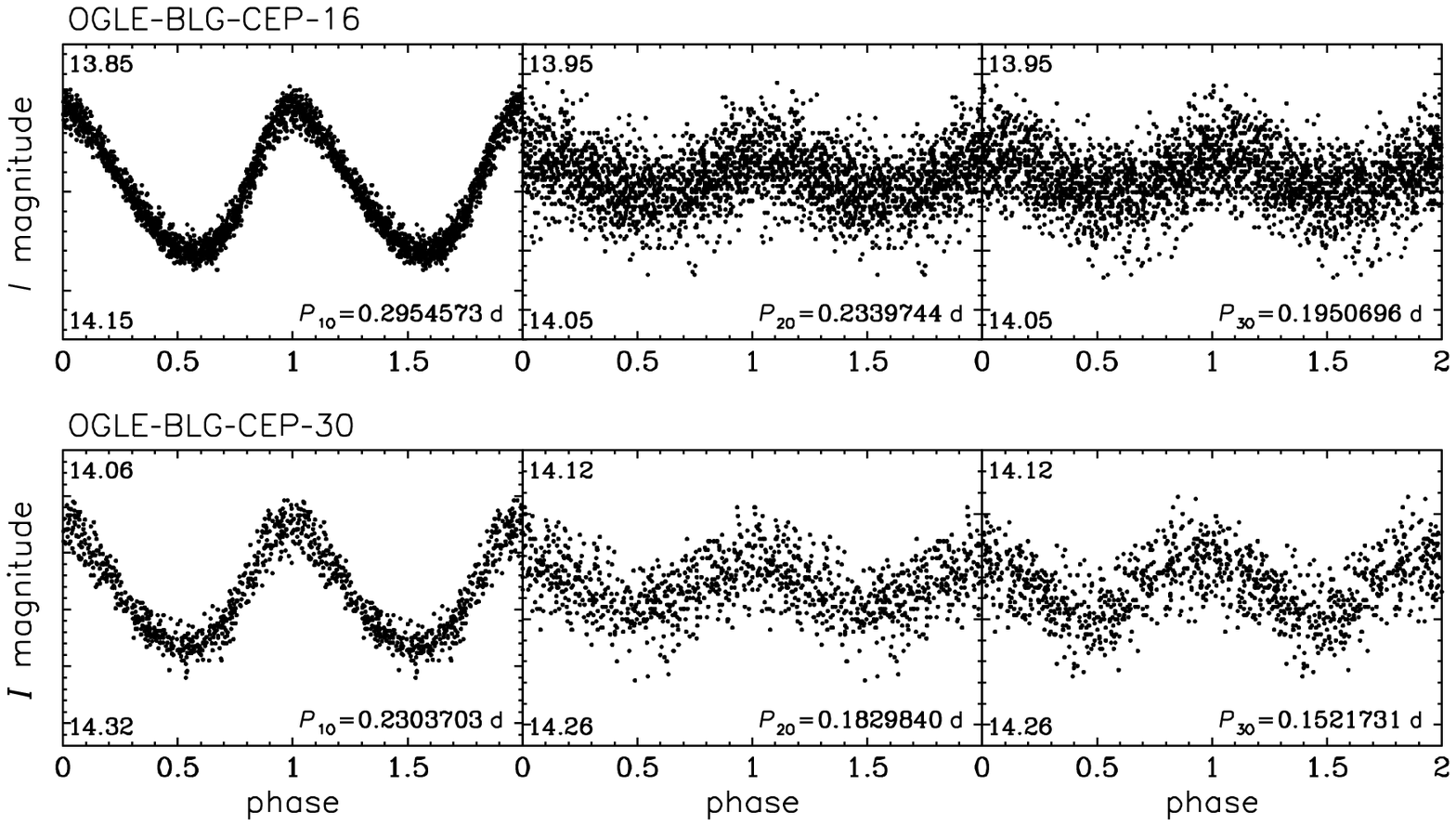}}
\FigCap{{\it I}-band light curves of triple-mode Cepheids in the Galactic
bulge. Each mode is shown after prewhitening with the other two modes. Note
that the range of magnitudes varies from panel to panel. Numbers in the
left corners show the minimum and maximum magnitudes of the range.}
\end{figure}

The Petersen diagram (shorter-to-longer period ratio \vs the logarithm of
the longer period) is shown in Fig.~9. We also plotted here double- and
triple-mode Cepheids from the Magellanic Clouds (Soszyñski \etal 2008a,
2010a) and $\delta$~Sct stars from the LMC (Poleski \etal 2010). The
diagram reveals another remarkable property of multi-mode Cepheids in the
bulge: the ratio of the second-overtone to first-overtone periods is
distinctly lower than in the Magellanic Clouds. Similar feature concerns
the period ratios of the third and first overtones, and third and second
overtones in triple-mode Cepheids.

\begin{figure}[htb]
\vglue-3mm
\centerline{\includegraphics[width=11.3cm, bb=45 270 555 755]{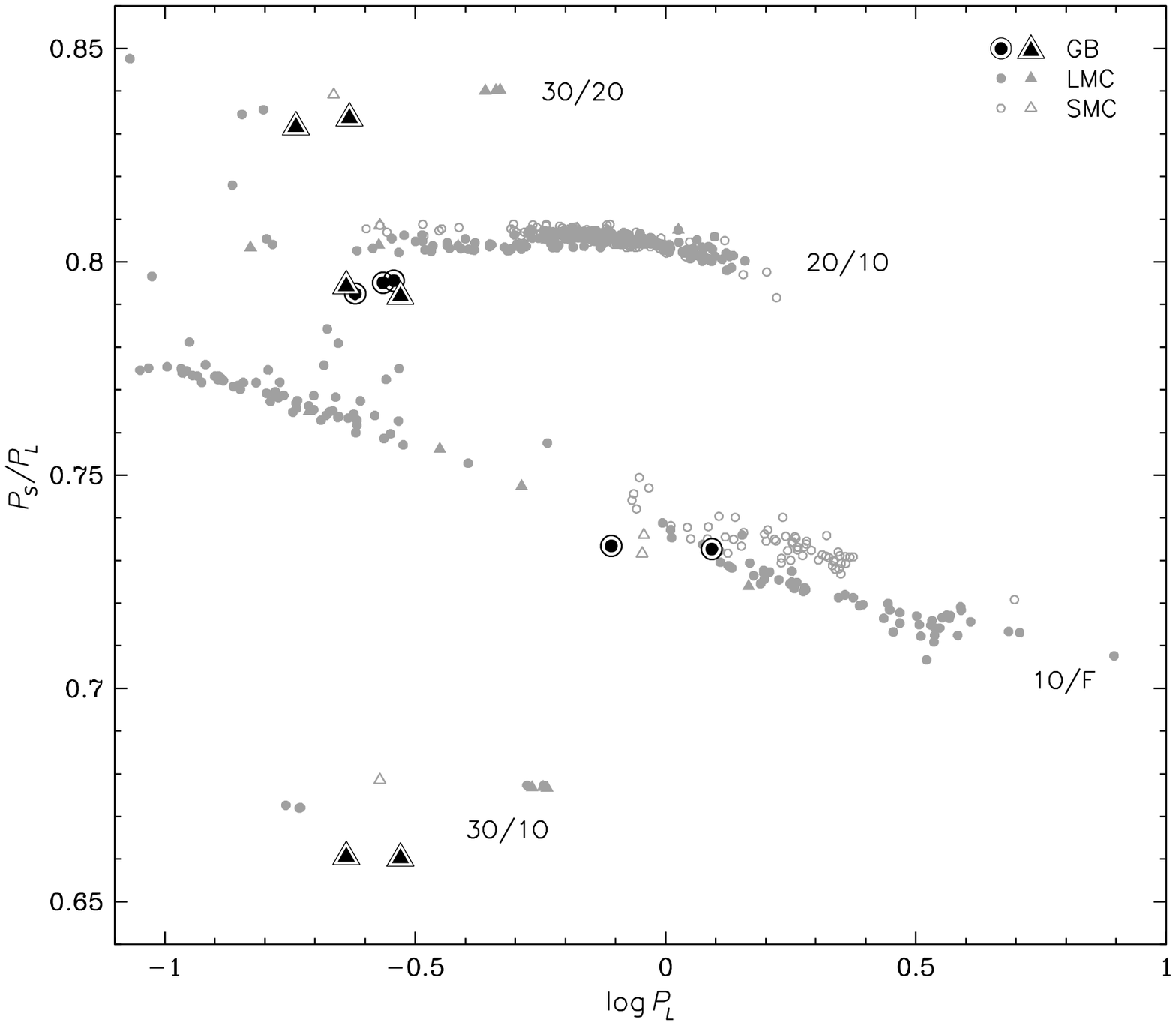}}
\FigCap{Petersen diagram for multiperiodic Cepheids in the Galactic bulge
(black symbols), LMC (grey solid symbols), and SMC (grey empty
symbols). Circles represent double-mode Cepheids (one point per star),
while triangles show the period ratios in triple-mode Cepheids (three
points per star).}
\end{figure}

\Subsection{Type~II Cepheids}
The PL diagram (Fig.~7) shows that most of type~II Cepheids in our sample
are placed in the Galactic bulge. The scatter of the relation can be
largely explained by the depth of the bulge along the line of sight. Six
type~II Cepheids (two BL~Her, one W~Vir and three RV~Tau stars) are by
about 3~mag fainter than type~II Cepheids of the same periods in the bulge,
and these are probably members of the Sgr dSph, like RR~Lyr stars also
visible in the lower part of the PL diagram.

Comparison to the PL relations followed by type~II Cepheids in the LMC
(Soszyñski \etal 2008b) reveals significant differences. In the LMC RV~Tau
stars are systematically brighter than the PL relation fitted to BL~Her and
W~Vir stars, while in the bulge such a behavior is not detectable. On the
other hand in the LMC (Soszyñski \etal 2008b) the PL relations of BL~Her
and W~Vir stars were co-linear within the uncertainties while in the bulge
BL~Her stars are distinctly brighter than the PL relation defined by W~Vir
stars. This feature is similar to that observed for type~II Cepheids in the
SMC (Matsunaga \etal 2011a). It is worth noting that RRab stars, also
plotted in Fig.~7, seem to be on the extension of the PL relation for
BL~Her stars. So, brighter BL~Her stars imply brighter RR~Lyr stars in the
Galactic bulge.

Fig.~2 presents examples of light curves of BL~Her, W~Vir, and RV~Tau stars
from our catalog. W~Vir stars have on average more scattered light curves
than BL~Her stars due to variations of period, phase and sometimes
amplitude. However, it seems that type~II Cepheids in the Galactic bulge
have more stable light curves than their counterparts in the Magellanic
Clouds. There are also exceptions to this rule, such as a W~Vir star
OGLE-BLG-T2CEP-059, with so rapid period changes that its OGLE light curve
cannot be phased using a constant period.

Our sample of type~II Cepheids contains objects of particular interest. We
discovered first two cases of BL~Her stars with the period doubling
(OGLE-BLG-T2CEP-257, OGLE-BLG-T2CEP-279), the behavior predicted
theoretically by\linebreak Buchler and Moskalik (1992). An in-depth
analysis of OGLE-BLG-T2CEP-279 (BLG184.7 133264) was performed by Smolec
\etal (2011).

\begin{figure}[htb]
\centerline{\includegraphics[width=12.2cm, bb=60 615 570 755]{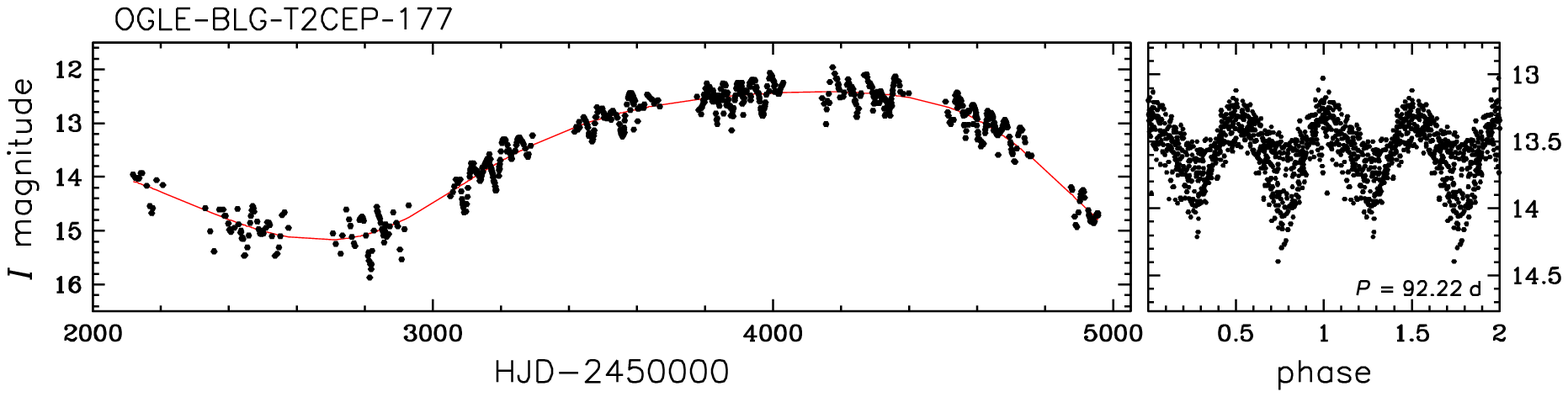}}
\FigCap{{\it I}-band light curve of an RVb star OGLE-BLG-T2CEP-177. {\it
Left panel} shows the unfolded light curve. Red line is the spline function
fitted to the light curve. {\it Right panel} presents the light curve folded
with the ``double'' pulsation periods after subtracting the long-term trend.}
\end{figure}

Fig.~10 presents the light curve of OGLE-BLG-T2CEP-177 -- an RV~Tau star
with variable mean magnitude (RVb type). If the changes of the mean
brightness are periodic (like in other RVb stars, \eg OGLE-LMC-T2CEP-200),
the eight years of the OGLE-III observations covered only one cycle of
these variations. It shows the advantage of the long-term OGLE survey. This
objects and other variable stars from this catalog are continuously
monitored during the present stage of the OGLE survey -- OGLE-IV.

\vspace{0.5cm}

\Acknow{We are grateful to W.A.~Dziembowski for fruitful comments and
suggestions. We want to thank Z.~Ko³aczkowski and A.~Schwarzenberg-Czerny
for providing software which enabled us to prepare this study.

The research leading to these results has received funding from the
European Research Council under the European Community's Seventh Framework
Program\-me (FP7/2007-2013)/ERC grant agreement no. 246678. This work has
been supported by the MNiSW grant no. IP2010 031570 (the Iuventus Plus
program) to P. Pietrukowicz. The massive period search was performed at the
Interdisciplinary Centre for Mathematical and Computational Modeling of
Warsaw University (ICM), project no. G32-3. We wish to thank M. Cytowski
for his skilled support.}

\end{document}